# Giant Vesicles with Encapsulated Magnetic Nanowires as Versatile Carriers, Transported via Rotating and Non-Homogeneous Magnetic Fields

**Mateos-Maroto Ana[1], Ortega Francisco[1,2], Rubio G. Ramón[1,2], Berret Jean-François[3], Martínez-Pedrero Fernando[1*]**

[1]*Departamento de Química-Física, Universidad Complutense de Madrid, Avda. Complutense s/n, Madrid, 28040, Spain.*
[2]*Universidad Complutense de Madrid, Inst. Pluridisciplinar, Paseo Juan 23,1, E-28040 Madrid, Spain.*
[3]*Laboratoire Matière et Systèmes Complexes, UMR 7057 CNRS Université Denis Diderot Paris-VII, Bâtiment Condorcet, 10 rue Alice Domon et Léonie Duquet, 75205 Paris, France*

**E-mail**: fernandm@ucm.es

**Abstract:** In this work, two real-time methods to transport magnetic nanowires confined in giant hybrid vesicles upon the application of different strategies are studied. The microscale carriers are either magnetically guided through the viscous medium by a non-homogenous field or advected by precisely monitored hydrodynamic flows. The slender geometry of the magnetic component enables the application of large torques, the in-situ characterization of the rotational dynamics, as well as the guided propulsion via the continuous thrust of the nanowire tip on the confining bilayer. The flexibility of the vesicles, required to deform along the steering direction during passage through microscale openings, is enhanced via the adsorption of non-ionic surfactants on the lipid membrane. The resulting integrated system is an excellent candidate to transport colloidal cargos or fluid amounts of some picoliters in microfluidic platforms, even in physiological environments, since it combines the maneuverability of propelled microscopic systems and the protection conferred by the vesicle.

## 1 - Introduction

The transport of micro/nano-cargos through a viscous fluid, at low Reynolds number, plays a key role in many different contexts of biology and technology. The different strategies employed to generate propulsion, for example the application of drag forces through external fields, the rotation of a chiral object in the medium or spheres in the proximity of an interface, the wriggling of flexible structures or the generation of hydrodynamic flows via surface-chemical reactions, are compatible with the negligible role of the inertial forces and overcome the restrictions imposed by the Scallop theorem, which states that the configurations sequentially adopted by a swimmer cannot be time-symmetric.[1-7]

In most of the practical applications, nanoformulations such as micelles, vesicles and nanoparticles are preferred since they are able to pass through physical and biological barriers.[8-13] However, the strong thermal fluctuations of the nano-transporters, only rectified into directed motion under the application of large external forces, strongly hinder the performance of planned trajectories.[14-15] Designs based on externally actuated microscale particles, with sizes in the same range as most eukaryotic cells, partially overcome the randomization generated by the thermal noise. These carriers are able to transport drugs, nanocargos, colloidal particles or biological entities through a viscous fluid via externally induced flows, subtle hydrodynamic interactions that minimize the damage suffered by the cargo during the transit. However, their relatively large size can be an inconvenient when transported through confined environments.

A potential solution that combines monitored transport, adaptability and protection, essential in many biological applications, is the development of membrane-based microscale carriers.[16] Hollow microobjects with mechanically deformable vesicles and no internal scaffolding, for instance macrophages, microorganisms, stem, sperm and red blood cells, or lipid vesicles, are among the most promising candidates to transport cargos through a viscous medium.[17-20] They are able to transit through tiny capillaries and narrow pores,[21-25] while the encapsulated content remains isolated from the potentially hazardous environment by the phospholipid bilayer, even in physiological conditions.

The aim of this work is to prepare a novel and simple hybrid platform as a vehicle for tailored transport in microfluidic conditions. In our design, giant vesicles consisting in a lipid membrane, fluid at room temperature, enclose a minute volume of aqueous solution together with one or more superparamagnetic nanowires.[26] These nanowires have been used in previous studies to transport colloidal cargo near a complex surface,[27-28] to introduce exogenous DNA into mammalian cells,[29] or as sensitive probes to determine viscoelastic properties of living cells[30-32] and monolayers of surface-active proteins.[33-36] By applying different magnetic field configurations, the vesicles are positioned by magnetic field driving or precisely propelled thanks to the induced rotation of the encapsulated nanorods. The slender geometry of the actuated wires enhances the magnitude of both the applied torque and the generated flow, as the same time as facilitates the precise monitorization of the angular dynamics. On the other hand, the use of nanowires, with a diameter of a few hundreds of nanometers, minimizes the size of the actuated component in one of the dimensions, which can be crucial in the transport of the carrier through confined geometries. With this purpose in mind, relatively inextensible bilayers are transformed into highly deformable membranes by incorporating a non-ionic surfactant with a lower packing parameter into the bilayer membrane.[37-38] Amphiphilic molecules change the dynamics and structure of the bilayer, accumulating at the most stressed sites in the membrane and lowering the energetic cost of its deformation.[24, 38-39] The mechanical deformable biohybrid capsules allow for the manipulation and transport of molecules in a protective environment, through entangled paths, without requiring direct contact or chemical binding.

## 2 – Results and Discussion

### 2.1 - Lipid vesicles containing magnetic nanowires

The application of the slightly modified electroswelling protocol described in the Experimental Section, where magnetic nanowires are dispersed in the aqueous solution before the application of the AC electric field, results in the formation of polydisperse uni- or multilamellar vesicles with sizes ranging from less than 1 to 30 μm. After sedimentation, the vesicles and non-encapsulated nanowires remain suspended above the glass surface of the measurement cell due to the balance between gravity and the electrostatic repulsion with the negatively charged glass substrate. The giant vesicles display Brownian diffusion on the plane parallel to the surface, the ($x,y$) plane, and negligible upright Brownian movement. A small proportion of the vesicles host in their internal aqueous pool a varied number of magnetic nanowires, which become entrapped during the electroformation process. Nanowires were fabricated through a co-assembly process assisted by the presence of an external magnetic field. The protocol for the synthesis of the wires is described in the Experimental Section and Reference [30]. Before encapsulating them inside the lipid vesicles, they were sonicated for several minutes to shorten their lengths. The resulting nanowires have an average length of 2.4 μm and a dispersity of 0.35. On the other hand, the size distribution of the vesicles was reported in Reference [40]. Even when most of the formed vesicles have sizes below 1 micrometer, being unable to host nanowires, the percentage of vesicles with diameters between 10 and 20 micrometers containing one or more nanowires is larger than 20%.

In **Figure 1** we show a set of microscopy images that illustrates the principal types of loaded vesicles. Wires with a length significantly smaller than their wrapping vesicles, present the same anisotropic diffusive motion in the ($x,y$) plane as the one exhibited by the non-encapsulated nanowires, Figure 1 a and b. A subdiffusive motion of the bounded wires, where their mean squared displacement is not linear with time, is detected only when the vesicles are densely populated, Figure 1 c, when the motion of the rod within the vesicles is restricted by the presence of internal vesicles, lipid residues or partitioning lipid bilayers, or when the length of the confined wires is comparable to the confining vesicle diameter, Figure 1 d, e and f. Occasionally, those nanowires slightly deform the lipid bilayer, Figure 1 d and e, or become bent during the encapsulation process, Figure 1 f.

## 2.2. Vesicles rotating parallel to the substrate

In the first series of experiments, we characterize the rotation of a couple of nanowires, $L$ = 11.1 μm and 3.0 μm in length and $D$ = 0.5 μm in diameter, confined in two lipid vesicles 11.2 μm and 13.1 μm in size respectively. In these vesicles, there is no contact between the inner lipid leaflet and the nanowires. The latter float at a certain distance from the membrane due to the balance between gravity and electrostatic interactions with the zwitterionic lipids. The spinning is induced by a circularly polarized rotating magnetic field with amplitude $H_0$ and angular frequency $\omega_{field}$ = $2\pi f_{field}$, applied on the substrate plane ($x,y$), $\vec{H}_{xy} = H_0\left(cos(\omega_{field}t), sin(\omega_{field}t), 0\right)$, see schematic in **Figure 2** a. The magnetic torque applied on the nanowires, $\vec{\tau}_m = \mu_0 V \langle \vec{m} \times \vec{H} \rangle = \frac{1}{2}\mu_0 V \Delta\chi H^2 sin(2\beta)\hat{i}$, is instantaneously balanced by the viscous torque $\vec{\tau}_v = -\left(\pi L^3/3g(L/D)\right)\eta_0\, d\theta/dt$ produced by the wire spinning in the continuous medium of viscosity $\eta_0$. In these equations, $V$ is the wire volume, $\vec{m}$ its magnetization, $\mu_0$ the permeability in vacuum, $\langle...\rangle$ denotes a time average, $\Delta\chi = \chi^2/(2+\chi) = 2.3\pm0.7$, $\chi = 3.6\pm0.9$ the magnetic susceptibility, $\beta$ the angle between the wire main axis and the applied field, and $g(L/D) = \ln(L/D) - 0.662 + 0.917(D/L) - 0.05(D/L)^2$ is a dimensionless parameter that only depends on the anisotropy ratio ($D/L$).[30] At relatively high field strength, green regions in Figure 2 a and b, the nanowires synchronously rotate with the external field at a constant angular velocity $\omega_{nanowire} = 2\pi f_{nanowire} = 2\pi f_{field}$. In this regime, the angle $\theta$ rotated by the nanowire linearly increases with time (green circles in Figure 2 c). The wire delays with respect to the rotating field, forming a constant angle $\beta$ such that the magnetic torque equals the viscous torque. As the strength of the applied field decreases $\beta$ increases, the magnetic torque remains constant and the wire continues rotating at the angular velocity imposed by the external field. However, below a critical field strength $H_c$, $\beta$ becomes larger than $\pi/4$ radians, and the magnetic torque turns out to be smaller than the drag torque exerted on the wire at the field angular frequency $\omega_{field}$, preventing the wire rotating synchronously with the field. In the new regime, the angle $\beta$ between the wire and the rotating field, the difference between the red circles and the green line in Figure 2 c, monotonously increases with time. When $\beta$ is between $\pi$ and $2\pi$ rad, the nanowire is forced to rotate in the contrary direction and the angle $\theta$ travelled by the rod decreases. Once $\beta$ reaches $2\pi$ rad the previous angular dynamics is repeated, as depicted by the periodic pattern described by the red circles in Figure 2 c. Hence, by decreasing the field strength below a critical value $H_c$, we observe the expected transition between a synchronous and an asynchronous regime (green and red zones respectively, in Figure 2 a and b).[41] For a slender cylinder with length $L$ and diameter $D$, the critical field strength $H_c = \sqrt{8\eta_0 L^2 \omega \big/ 3\mu_0 \Delta\chi D^2 g\left(\dfrac{L}{D}\right)}$ is proportional to $\omega^{1/2}$. The continues lines in Figure 2 a and b, are the corresponding fits to the measured values, with an effective viscosity of the water environment of $(2.5\pm0.3)\times10^{-3}$ and $(2.1\pm0.3)\times10^{-3}$ Pa s respectively. The expected value of the viscosity for the 0.3 M sucrose solution internal pool is $1.2\times10^{-3}$ Pa s,[42] and the increase in this value can be attributed to the proximity of the solid interface and membrane. At lower field strength, the magnetic torque becomes unable to induce any rotation on the wire, gray zones in Figure 2 a and b, and grey squares in Figure 2 c.

To characterize the effect that the nanowire rotation has on the vesicles, we take advantage of the occasional presence of vesicles consisting of a small but perceptible vesicle adhered to a larger one, which in turn contains a nanowire, **Figure 3** and **Movie 1**. The induced spinning and revolving motions of the encapsulated magnetic nanowires generate an internal flow that causes the spinning of the vesicles at a slower rate in the direction imposed by the field. In Figure 3 a, we show how a rotating nanowire, 10.5 µm in length and encapsulated in a vesicle 11.0 µm in size, passes from an asynchronous to a synchronous regime above a critical field $\mu_0 H_c = 3.2$ mT, and how the induced rotation of the vesicle presents a scaled behavior: below $H_c$ the rotational frequency of the vesicle increases with the field strength, whereas above $H_c$ it remains roughly constant. The linear response of the membrane to the wire actuation is confirmed by Figure 3 b, which shows that in the asynchronous regime the temporal response of the vesicle also presents a back-and-forth rotation, and that the angles covered by the nanowire and membrane almost overlap when the former is multiplied by $\varepsilon$, a nondimensional parameter determined by the efficiency of the momentum transfer mechanism and assessed as the ratio between the vesicle and nanowire rotational frequency. As expected, inertia has not any effect on the studied dynamic at low Reynolds number and the membrane instantaneously responds to the flows generated by the rotating nanowire. The time evolution of the angle travelled by the rotating wire and vesicle in the asynchronous regime is given by:[43]

$$\theta_{nanowire}(t) = 2\pi f_{field} t - \arctan\left[ f_c/f_{field} + \sqrt{1-\left(f_c/f_{field}\right)^2} \tan\left(\left(2\pi f_{field}\sqrt{1-\left(f_c/f_{field}\right)^2}\right)t\right) \right] \quad (1)$$

, and

$$\theta_{vesicle}(t) = \varepsilon\left( 2\pi f_{field} t - \arctan\left[ f_c/f_{field} + \sqrt{1-\left(f_c/f_{field}\right)^2} \tan\left(\left(2\pi f_{field}\sqrt{1-\left(f_c/f_{field}\right)^2}\right)t\right) \right] \right). \quad (2)$$

In the curves shown in Figure 3 b, the fitting parameter is $f_c$ =17.8 Hz, comparable to the value detected in Figure 3 a. Even when the theoretical curve 2 only fits the measured angle rotated by the vesicle in limited time windows, it properly predicts the ranges of frequency at which the angle travelled by the vesicle decreases. Here, the slight differences between the experimental data and the theoretical predictions can be attributed to the hydrodynamic interactions with other objects placed out of the confining membrane. In Figures 3 c (i-ii), the synchronous/asynchronous transition is induced by increasing the field frequency at two different values of the field strength. In the synchronous regime, the angular velocity of both the nanowire and actuated vesicle increases linearly with the field frequency, but at field frequencies higher than $f_c = 3\mu_0 \Delta\chi D^2 g\left(\frac{L}{D}\right) H_c^2 / \left(16\pi\eta_0 L^2\right)$, the nanowire enters in the asynchronous regime, where their average rotational frequency is given by $f_{nanowire}(f) = f_{field} - \sqrt{f_{field}^2 - f_c^2}$.[31] The lipid membrane responds linearly to the external actuation, and above $f_c$ the average rotational frequency of the vesicle is fitted by the scaled function $f_{vesicle}(f) = \varepsilon\left(f_{field} - \sqrt{f_{field}^2 - f_c^2}\right)$. For the hybrid vesicle of Figure 3, $\varepsilon$ = 0.04 in the range of frequencies and field strengths applied.

We have demonstrated how the induced spinning and revolving motions of the nanowires within the vesicles generate a flow field that causes the spinning of the vesicles in the same direction. By following the trajectory of small tracers, liposomes or phospholipid residues leftover from the synthesis, we prove in **Figure 4** that the rotation of the giant vesicle also perturbs the fluid outside. When the hybrid motors rotate around the axis perpendicular to the substrate, they generate a rotating flow field $\vec{v}$ that decreases with distance to the vesicle center as $r^{(-2.4\pm0.3)}$, as shown in Figure 4b. Since a sphere rotating at a constant angular velocity generates a flow field that decay asymptotically as $r^{-3}$, the observed difference must be due to finite size effects and to the presence of the solid substrate.[44] Hence, the generated flow may be used to manipulate particles in the viscous medium, rotating them around the actuated vesicle in the direction imposed by the external field, or to achieve an effective mixing of fluids.[45]

### 2.3. Hybrid vesicles transported under the action of a rotating magnetic field

The flow generated by the monitored rotation of the bounded nanowires can be used to propel the hybrid system near a solid boundary. When the vesicle rotates around an axis perpendicular to the substrate, the flow generated and rectified by the solid surface has no preferential direction in this plane and the rotating vesicle is not advected on the substrate. However, if the rotation of the vesicle is induced by a circularly polarized magnetic field rotating in the plane perpendicular to the glass surface $\vec{H}_{yz} = H_0\left(0, cos(\omega t), sin(\omega t)\right)$, see schematic in **Figure 5**, then the spinning motion of the hybrid vesicles is rectified into a slipping/rolling motion due to hydrodynamic coupling with the glass surface, see **Movie 2**. The hydrodynamic conveyor belt generated by the rotating hybrid vesicle in the presence of the external boundary, a cooperative flow that can be used as an efficient mechanism to transport matter in the low Reynolds number regime,[40] produces a net thrust that advects the micromotor in the direction perpendicular to the rotation axis. In Figure 5 we show the average speed along the $x$ axis, $v_x$, of an isolated hybrid vesicle containing a rotating nanowire. Data are displayed versus the rotating field strength, Figure 5 b, and frequency, Figure 5 c, and compared to those obtained from the freely dispersed nanowires, $v_{nanowire}$. At constant frequency, Figure 5 b, the measured linear velocities qualitatively reproduce the trends showed by the angular velocity in Figure 3 a: the linear velocity of the propelled vesicles increases with the field strength in the asynchronous regime, while it remains constant once the field strength is higher than $H_c$. Non-encapsulated nanowires having a similar size present an analogous behavior, but with larger velocities and critical field strengths. Unconfined nanowires move at larger linear velocities because of their lower drag coefficients, and probably due to the use of a more efficient mechanism of momentum transfer through the viscous liquid. On the other hand, the enhancement observed in the critical field value reveals that the confinement increases the viscous torque applied on the rotating nanowires. At constant field strength, Figure 5 c, the linear velocity of the hybrid vesicles grows linearly below the critical frequency $f_c$, while above decreases as $v_{hybrid\ vesicle}(f) = D_{vesicle} \gamma \varepsilon \left( f_{field} - \sqrt{f_{field}^2 - f_c^2} \right)$. Here, $f_c$ = 50 Hz, the vesicle diameter $D_{vesicle}$ = 11.2 μm, and $\gamma$ = 0.448 is a non-dimensional parameter included to consider that the vesicle presents a rolling and slipping motion. According to the previous results, the velocity is maximized at intermediate frequencies, when the rotational dynamics of the confined nanowire and membrane changes from an asynchronous to a synchronous regime. In this range, the lipid vesicles are transported at mean speeds of up to 5 micrometers/second, compatible with the swimming velocities of other membrane-based driven microswimmers.[18,46-47] In this experimental configuration, the direction of motion can be changed without difficulty, as shown in other analogous systems, by substituting the $x$ field by a rotating field applied along the $y$ direction.[5,7] To conclude this part, it is interesting to point out that the hydrodynamic flows generated by the nanorods are larger than those observed when the flow is generated by the induced rotation of spherical superparamagnetic particles.[40] Consequently, the velocity of transport of the giant vesicles increases by an order of magnitude.

**2.4. Hybrid vesicles transported under the action of a magnetic field gradient**
An alternative procedure for propelling the loaded vesicles through the viscous medium is to position them by the action of magnetic field gradients, a technique named magnetic tweezers.[48] A spatial inhomogeneity in the applied field causes forces on the nanowires, that move in the direction where the field strength is highest.[49] The field gradient persistently pulls nanowires contained into the vesicles, with their main axes aligned along the magnetic field direction. By using the magnet configuration described in the Experimental Section, vesicles 10 μm in size reach speeds in the order of 1 μm s$^{-1}$, which in the stokes regime and in the proximity of the rigid wall correspond to magnetic forces around 0.1 pN.[50] The direction of motion is changed by displacing the magnets positioned below the sample, in a configuration such that produces a magnetic field parallel to the glass substrate and avoid the occurrence of rising magnetic forces. By placing the magnet on top of the measurement cell, the vesicles are dragged upward and separated from the confining plane.

In the next series of experiments, we show how the hybrid vesicles can be steered through different microfluidic constrictions by employing the field gradient method. Here, the adaptability of the vesicles is increased by adding 0.38 mM Triton X-100 in the aqueous phase, a widely used non-ionic surfactant for recovery of membrane components under mild non-denaturing conditions. The inclusion of a small amount of surfactant in the lipid bilayer results in the display of larger membrane vibrations, clearly appreciable in the bright-field microscope, and in alterations in the shape of the vesicles, which lose their tendency to adopt a spherical symmetry, **Figure 6**, image I. On the other hand, the presence of the detergent also reduces the line tension, favoring the formation of transient pores that close when the internal content leaks out. This may limit the use of the vesicles as transporters.[51-52] In that case, vesicles of lipid membranes in the liquid-ordered phase with encapsulated magnetic nanowires can be a more stable alternative, since they strongly reduce the leakage induced by the presence of the detergent.[53] A higher surfactant concentration causes lipid solubilization, structural instabilities and disintegration of the bilayer.[54]

The carrier's maneuverability is shown in Figure 6 a, **Movie 3**, where a deformable hybrid vesicle is dynamically guided towards the south (I-V) and north-west (V-IX) of the observation area. The vesicle moves at a constant velocity of 1.1 μm s$^{-1}$ and forced to pass through two gaps of 5.6 and 3.7 microns, between immobile 10 μm polystyrene particles firmly attached to the substrate. The explored path is tuned by adjusting on the fly the orientation and direction of the driving field. To ensure that the steered vesicles do not use the space over the particles as way of escape we adopt two experimental cautions: the guiding magnet is placed below the sample, in a configuration such that produces a magnetic force pointing to the glass substrate, and we select vesicles with diameters close to the size of the obstacles. Besides, processes where the vesicles get out of the microscope focus were systematically neglected.

To quantify the deformation of the vesicle when it passes through an aperture, we have used an ImageJ plugin that fits an egg-like-geometry to the bright field microscope images.[55] During the transport, the membrane adopts an egg-like-shape in order to relax the stress exerted by the steering wire on the leading edge. The curve of an ellipse weighted by a normal distribution, $r = ae^{(-l^2+2cl-c^2)/(2b^2)}\sqrt{l-l^2}/\pi b$, is fitted to a hand-picked set of points of the membrane outline, as shown in Figure 6, image III. Here, $l$ is the coordinate along the egg's long axis, $r$ is the corresponding perpendicular distance to the membrane, $a$ is the width of the egg, $b$ the standard deviation of the distribution and $c$ the position of the distribution's peak along the length. The average deviation to an ellipse is given by $ellipse\ deviation = \sum_{l=0}^{l=1}\left(\frac{a}{2}\left(2\sqrt{l-l^2}\right)-r\right)^2/n$, where $n$ is the number of points measured along the egg's length. The steered vesicle decelerates from 1.1 to 0.3 μm s$^{-1}$ before passing between two obstacles, Figure 6 a and b, III, and the previous image analysis reveals that the vesicle adopts a more elliptic geometry and a minimum width when passing through the opening. While the deformable vesicle passes without difficulty through the 5.6 μm aperture, it spends almost a minute to go across the narrowest one. In fact, in our experimental set-up we never detected that the vesicles were able to pass through gaps thinner than the 20% of the vesicle diameter. Even when the increased deformability, qualitatively noticeable by comparing the vesicles shown in Figures 1-5 to those depicted in Figure 6 (Movie 3), allows for an optimized steered exploration of disorder and confining environments, the applied magnetic forces, in the order of 0.1 pN, are not strong enough to provoke remarkable extra-deformations during the passage.

In an attempt to enhance the deformability of the membranes, we also increased the temperature of the vesicles composed by pure DOPC membranes from 25 ºC to 70 ºC, but this change did not significantly alter the stability of the lipid bilayers or the capability of the vesicles to circulate through narrow apertures. Even when magnetic carbon nanotubes were employed in a previous work for spearing cell membranes,[29] we have never detected that the magnetic forces here employed were sufficiently strong to pierce and go through the lipid bilayer.[56] Besides the basic geometries shown in Fig. 6, circuits having more complex patterns can be equally explored by the hybrid deformable capsules. The ability to transport hybrid vesicles close to a confining surface creates many opportunities to integrate them in a microfluidic platform.

# 3 - Conclusion

Through a slight modification of the classical electroformation method, dispersing the magnetic nanowires in the aqueous solution before the application of the AC electric field, we have formulated a new platform for the transport of cargos at low Reynolds number. Lipid vesicles containing one or more magnetic nanowires are transported through the implementation of different transport strategies: either the application of magnetic forces or the monitored advection of the hybrid structures. The combination of these two methods allows for a subtle and accurate control in the transport of the hybrid vesicles, both in bulk and in confined environments.[57] The versatility of the vesicles, their capability to be efficiently transported by means of both techniques, can be essential in conditions where it is impracticable to place a strong magnet in the proximity of the sample.

In the first approach, the fluid membranes respond linearly to the synchronous or asynchronous spinning motion of the confined nanowires, perturbing at the same time the fluid outside in a controlled manner. The field induced rotation is rectified into translational motion only in the proximity of an interface, condition that is fulfilled in most of the systems of application.[58-59] The vesicles reach velocities larger than the ones measured when they are loaded with superparamagnetic spherical particles. In superparamagnetic beads the susceptibility at static field is isotropic, and the induced torque is mainly determined at low frequency by the possible existence of a permanent moments in the particle, and at high frequency by the complex susceptibility.[60] Superparamagnetic nanowires, for which the applied torque is proportional to the susceptibility anisotropy $\Delta\chi$, enable the application of larger torques at moderate frequencies that results in enhanced propulsion velocities. In the second approach, the loaded lipid vesicles are steered through different microfluidic constrictions by employing the field gradient method.[61] Here, the inclusion of surfactants in the lipid bilayer increases the deformability of the vesicles, that display an enhanced ability to pass through narrow openings and confined geometries.

These vesicles can be disintegrated upon application of a NIR femtosecond laser pulse,[40] and used for the sheltered on-command release of picoliter droplets and chemicals through small passageways and pores, in physiological conditions or in lab-on-a-chip devices. By further studying custom-decorated deformable vesicles, we may be able to design reconfigurable drug delivery systems for diagnostic or therapeutic use. These cell-like compartments, containing wires that present a strong and measurable response to the presence of constant, rotating and non-homogenous magnetic fields, can be also employed as microreactors and micromixers,[61-64] as model systems in the induced diffusion through membranes,[65] the study of the mechanical properties of the lipid bilayers,[66] as well as for understanding the transport and extravasation of biological cells.

# 4 – Experimental section

*Magnetic Wire Synthesis and Characterization*
Nanowires were fabricated through a bottom-up co-assembly process using superparamagnetic $\gamma$-Fe2O3-coated particles and oppositely charged polyelectrolytes poly(trimethylammoniumethylacrylate)-*b*-poly(acrylamide) with molar mass of 11,000 g mol$^{-1}$ for the cationic block and 30,000 g mol$^{-1}$ for the neutral block. Obtained from transmission electron microscopy (Jeol-100 CX), the $\gamma$-Fe2O3 size distribution was characterized by a median diameter of 13.2 nm and a dispersity of 0.23. The dispersity is here defined as the ratio between standard deviation and average diameter. Reference [30] illustrates in more detail the protocol for the fabrication of the wires. Following the described protocol, the as-prepared samples contained anisotropic objects of median length 12.2 µm and of median diameter 0.49 µm. The wire geometrical characteristics were determined from optical microscopy using an Olympus x100 objective of spatial resolution 0.28 µm and from scanning electron microscopy (Supra 40V SEM-FEG, Zeiss). The dispersities in length and diameter were found to be 0.50 and 0.20 respectively. Wire lengths were finally shortened to an average length of 2.4 µm and a dispersity of 0.35 by sonicating the samples for 5 minutes (W = 50 watts). To prevent bacterial contamination, the samples were autoclaved at 120 °C and atmospheric pressure for 20 minutes (Tuttnauer Steam Sterilizer 2340M), concentrated by magnetic sedimentation at $10^6$ wires µl$^{-1}$ and stored at 4 °C.

*Encapsulation*
The encapsulation of the paramagnetic nanowires in giant vesicles made up of 1,2-dioleoyl-*sn*-glycero-3-phosphocholine (DOPC), purchased from Avanti Polar Lipids (Alabaster, AL), was achieved by adding an aliquot of an aqueous stock solution of paramagnetic wires to a sucrose solution during the classical electroformation method.[67] 250 µL sample of a solution (1 mg mL$^{-1}$) obtained by dissolving the lipid in chloroform was deposited by drop-casting on the conducting side of two glass slides, thoroughly washed with ethanol, distilled water, ethanol and dried before use, and coated with indium tin oxide (ITO, Sigma Aldrich, resistance 15-25 Ω/sq). Once the solvent was evaporated, we fabricated a closed chamber by separating these electrodes with a 2.0 mm silicone isolator (Himatra). Finally, the chamber was filled with a sucrose aqueous solution (0.3 M) and an AC electric field was applied to each ITO glass plate in three steps, following the protocol described in Reference [68]. In the first step, we applied a sinusoidal electric field, with a function generator (Agilent 33120A, Agilent Technology), at a frequency of 8 Hz and initial amplitude of 10 mV, which was gradually increased by 25 mV every 5 min for 1 h. In the second step, we kept the amplitude constant at 300 mV for 3 h. Finally, we applied a square field with a frequency of 4 Hz and amplitude of 300 mV for 1 h. This procedure led to the formation of uni- and multilamellar GVs with a size distribution range of 0.5–50 µm, some of them hosting a varied number of superparamagnetic nanowires in their internal aqueous pool. The membrane of the vesicles, composed of phospholipids having two unsaturated fatty acid chains, is fluid at room temperature. The nanowires within the vesicles, confined by the lipid bilayers in a small volume of aqueous sucrose solution, are not able to cross the membrane. To facilitate the use of optical transmission microscopy and favoring the sedimentation toward the substrate by gravity, the vesicles and magnetic nanowires were dispersed in a glucose solution (0.3 M), with lower density and a different refractive index.[69] The deformability of the membranes increases in the presence of a non-ionic surfactant, Triton X-100 (Sigma Aldrich), which interacts with the membrane altering the membrane composition.

*Measurement Cell*
The obtained suspension was introduced by capillarity between a glass slide (RS France) and a microscope coverslip (Menzel-Glaser), both spaced by a double-faced adhesive tape and sealed with silicone oil 200 mPa (Sigma Aldrich), minimizing occasional convective motions. Because of density mismatch, the magnetic nanowires and vesicles settled on the glass surface, where remained suspended above the glass slide due to the balance between gravity and the electrostatic interactions with the negatively charged substrate. The sedimented giant vesicles display membrane vibrations and Brownian motion on the glass substrate, and negligible out-of-plane

Brownian displacement. Most of the encapsulated wires also present a diffusion movement characterized by anisotropic diffusion coefficients. All experiments were performed at room temperature.

*Magnetic Field and Monitorization*

To introduce different elliptically polarized rotating magnetic fields in the desired plane, $\vec{H}(t) \equiv H_1 cos(\omega t)\hat{i}_1 + H_2 sin(\omega t)\hat{i}_2$, being $\hat{i}_1$ and $\hat{i}_2$ two orthogonal unit vectors and $\omega$ the field frequency, the measurement cell was placed in the center of a custom-made system composed by two orthogonal pairs of coils, arranged on the stage of a bright-field microscope and aligned along the $x$ and $y$ axes. A fifth coil was located under the sample cell and aligned along the optical axis ($z$ axis). The coils were connected to three power amplifiers (Velleman k8060) commanded by a waveform generator (National Instruments 9269). In the application of a magnetic field gradient, a micropositioner (Narishige, modelo MMN-333) holds a custom-made rotational stage with a Neodymium-Iron-Boron permanent magnet 6 mm in. The magnet is positioned below the sample, at 2.0 mm in all the experiments, in a configuration such that produces a magnetic field parallel to the glass substrate and avoid the occurrence of lift forces. The particle dynamics was visualized with an upright optical microscope (BH2, Olympus) connected to a CCD camera (EO-1312M, Edmund) equipped with a $40 \times 0.65$ NA objective.

**Supporting Information**

As Supporting Information, we provide one .pdf file and 3 videoclips as support of Figures 3, 5a and 6a.

**Acknowledgements**

This work was supported in part by MINECO under the grant CTQ2016-78895-R. J.F.B acknowledges financial support from the ANR (Agence Nationale de la Recherche) and CGI (Commissariat à l'Investissement d'Avenir) through Labex SEAM (Science and Engineering for Advanced Materials and devices) ANR 11 LABX 086, ANR 11 IDEX 05 02 and through the contracts ANR-13-BS08-0015 (PANORAMA), ANR-12-CHEX-0011 (PULMONANO), ANR-15-CE18-0024-01 (ICONS), ANR-17-CE09-0017 (AlveolusMimics). F.M.-P. and A.M.-M acknowledge support from MINECO (Grant No. RYC-2015-18495).

# References

1. Purcell, E. M., Life at low Reynolds number. *American Journal of Physics* **1977,** *45* (1), 3-11.
2. Peyer, K. E.; Zhang, L.; Nelson, B. J., Bio-inspired magnetic swimming microrobots for biomedical applications. *Nanoscale* **2013,** *5* (4), 1259-1272.
3. Medina-Sánchez, M.; Xu, H.; Schmidt, O. G., Micro- and nano-motors: the new generation of drug carriers. *Therapeutic Delivery* **2018,** *9* (4), 303-316.
4. Sanchez, S.; Solovev, A. A.; Harazim, S. M.; Schmidt, O. G., Microbots Swimming in the Flowing Streams of Microfluidic Channels. *Journal of the American Chemical Society* **2011,** *133* (4), 701-703.
5. Martinez-Pedrero, F.; Massana-Cid, H.; Tierno, P., Assembly and Transport of Microscopic Cargos via Reconfigurable Photoactivated Magnetic Microdockers. *Small* **2017,** *13* (18), 1603449-n/a.
6. Massana-Cid, H.; Martinez-Pedrero, F.; Navarro-Agermi, E.; Pagonabarraga, I.; Tierno, P., Propulsion and hydrodynamic particle transport of magnetically twisted colloidal ribbons. *New J. Phys.* **2017**.
7. Martinez-Pedrero, F.; Tierno, P., Magnetic Propulsion of Self-Assembled Colloidal Carpets: Efficient Cargo Transport via a Conveyor-Belt Effect. *Phys. Rev. Appl.* **2015,** *3* (5), 6.

8. Gaumet, M.; Vargas, A.; Gurny, R.; Delie, F., Nanoparticles for drug delivery: The need for precision in reporting particle size parameters. *European Journal of Pharmaceutics and Biopharmaceutics* **2008**, *69* (1), 1-9.
9. Wilhelm, S.; Tavares, A. J.; Dai, Q.; Ohta, S.; Audet, J.; Dvorak, H. F.; Chan, W. C. W., Analysis of nanoparticle delivery to tumours. *Nature Reviews Materials* **2016**, *1*, 16014.
10. Guzmán, E.; Mateos-Maroto, A.; Ruano, M.; Ortega, F.; Rubio, R. G., Layer-by-Layer polyelectrolyte assemblies for encapsulation and release of active compounds. *Advances in colloid and interface science* **2017**, *249*, 290-307.
11. Ding, C.; Li, Z., A review of drug release mechanisms from nanocarrier systems. *Materials Science and Engineering: C* **2017**, *76*, 1440-1453.
12. Zhang, F.; Zhao, L.; Wang, S.; Yang, J.; Lu, G.; Luo, N.; Gao, X.; Ma, G.; Xie, H.-Y.; Wei, W., Construction of a Biomimetic Magnetosome and Its Application as a SiRNA Carrier for High-Performance Anticancer Therapy. *Advanced Functional Materials* **2018**, *28* (1), 1703326.
13. Yang, K.; Liu, Y.; Liu, Y.; Zhang, Q.; Kong, C.; Yi, C.; Zhou, Z.; Wang, Z.; Zhang, G.; Zhang, Y.; Khashab, N. M.; Chen, X.; Nie, Z., Cooperative Assembly of Magneto-Nanovesicles with Tunable Wall Thickness and Permeability for MRI-Guided Drug Delivery. *Journal of the American Chemical Society* **2018**, *140* (13), 4666-4677.
14. Yellen, B. B.; Hovorka, O.; Friedman, G., Arranging matter by magnetic nanoparticle assemblers. *Proc. Natl. Acad. Sci. U. S. A.* **2005**, *102* (25), 8860-8864.
15. Stoop, R. L.; Straube, A. V.; Tierno, P., Enhancing Nanoparticle Diffusion on a Unidirectional Domain Wall Magnetic Ratchet. *Nano letters* **2019**, *19* (1), 433-440.
16. Martínez-Pedrero, F.; Tierno, P., Advances in colloidal manipulation and transport via hydrodynamic interactions. *J. Colloid Interface Sci.* **2018**.
17. Tan S, W. T, Zhang D, Zhang Z., Cell or Cell Membrane-Based Drug Delivery Systems. *Theranostics* **2015**, *5* (8), 863-881.
18. Park, B.-W.; Zhuang, J.; Yasa, O.; Sitti, M., Multifunctional Bacteria-Driven Microswimmers for Targeted Active Drug Delivery. *ACS Nano* **2017**, *11* (9), 8910-8923.
19. Stanton, M. M.; Park, B.-W.; Vilela, D.; Bente, K.; Faivre, D.; Sitti, M.; Sánchez, S., Magnetotactic Bacteria Powered Biohybrids Target E. coli Biofilms. *ACS Nano* **2017**, *11* (10), 9968-9978.
20. Xu, H.; Medina-Sánchez, M.; Magdanz, V.; Schwarz, L.; Hebenstreit, F.; Schmidt, O. G., Sperm-Hybrid Micromotor for Targeted Drug Delivery. *ACS Nano* **2018**, *12* (1), 327-337.
21. Tannock, I. F.; Lee, C. M.; Tunggal, J. K.; Cowan, D. S. M.; Egorin, M. J., Limited Penetration of Anticancer Drugs through Tumor Tissue. *Clinical Cancer Research* **2002**, *8* (3), 878.
22. Chen, L. Microfluidics Synthesis, Characterization, and Applications of Bionspired Deformable Microparticles. Massachusetts Institute of Technology, Massachusetts (U.S.A.), 2018.
23. Chen, L.; An, H. Z.; Doyle, P. S., Synthesis of Nonspherical Microcapsules through Controlled Polyelectrolyte Coating of Hydrogel Templates. *Langmuir : the ACS journal of surfaces and colloids* **2015**, *31* (33), 9228-9235.
24. Romero, E. L.; Morilla, M. J., Highly deformable and highly fluid vesicles as potential drug delivery systems: theoretical and practical considerations. *International journal of nanomedicine* **2013**, *8*, 3171-3186.
25. Merkel, T. J.; Jones, S. W.; Herlihy, K. P.; Kersey, F. R.; Shields, A. R.; Napier, M.; Luft, J. C.; Wu, H.; Zamboni, W. C.; Wang, A. Z.; Bear, J. E.; DeSimone, J. M., Using mechanobiological mimicry of red blood cells to extend circulation times of hydrogel microparticles. *Proceedings of the National Academy of Sciences* **2011**, *108* (2), 586-591.
26. Luzzati, V.; Husson, F., The structure of the liquid-crystalline phasis of lipid-water systems. *The Journal of cell biology* **1962**, *12* (2), 207-219.
27. Zhang, L.; Petit, T.; Lu, Y.; Kratochvil, B. E.; Peyer, K. E.; Pei, R.; Lou, J.; Nelson, B. J., Controlled Propulsion and Cargo Transport of Rotating Nickel Nanowires near a Patterned Solid Surface. *ACS Nano* **2010**, *4* (10), 6228-6234.
28. Petit, T.; Zhang, L.; Peyer, K. E.; Kratochvil, B. E.; Nelson, B. J., Selective Trapping and Manipulation of Microscale Objects Using Mobile Microvortices. *Nano letters* **2012**, *12* (1), 156-160.

29. Cai, D.; Mataraza, J. M.; Qin, Z.-H.; Huang, Z.; Huang, J.; Chiles, T. C.; Carnahan, D.; Kempa, K.; Ren, Z., Highly efficient molecular delivery into mammalian cells using carbon nanotube spearing. *Nat Meth* **2005,** *2* (6), 449-454.
30. Berret, J. F., Local viscoelasticity of living cells measured by rotational magnetic spectroscopy. *Nat. Commun.* **2016,** *7*, 10134.
31. Chevry, L.; Colin, R.; Abou, B.; Berret, J.-F., Intracellular micro-rheology probed by micron-sized wires. *Biomaterials* **2013,** *34* (27), 6299-6305.
32. Celedon, A.; Hale, Christopher M.; Wirtz, D., Magnetic Manipulation of Nanorods in the Nucleus of Living Cells. *Biophysical Journal* **2011,** *101* (8), 1880-1886.
33. Dhar, P.; Cao, Y. Y.; Fischer, T. M.; Zasadzinski, J. A., Active Interfacial Shear Microrheology of Aging Protein Films. *Physical review letters* **2010,** *104* (1), 4.
34. Martínez-Pedrero, F.; Tajuelo, J.; Sánchez-Puga, P.; Chulia-Jordan, R.; Ortega, F.; Rubio, M. A.; Rubio, R. G., Linear shear rheology of aging β-casein films adsorbing at the air/water interface. *J. Colloid Interface Sci.* **2018,** *511*, 12-20.
35. Tajuelo, J.; Pastor, J. M.; Rubio, M. A., A magnetic rod interfacial shear rheometer driven by a mobile magnetic trap. *Journal of Rheology* **2016,** *60* (6), 1095-1113.
36. Tajuelo, J.; Pastor, J. M.; Martinez-Pedrero, F.; Vazquez, M.; Ortega, F.; Rubio, R. G.; Rubio, M. A., Magnetic Microwire Probes for the Magnetic Rod Interfacial Stress Rheometer. *Langmuir : the ACS journal of surfaces and colloids* **2015,** *31* (4), 1410-1420.
37. Bhatia, T.; Agudo-Canalejo, J.; Dimova, R.; Lipowsky, R., Membrane Nanotubes Increase the Robustness of Giant Vesicles. *ACS Nano* **2018,** *12* (5), 4478-4485.
38. Dimova, R.; Aranda, S.; Bezlyepkina, N.; Nikolov, V.; Riske, K. A.; Lipowsky, R., A practical guide to giant vesicles. Probing the membrane nanoregime via optical microscopy. *Journal of Physics: Condensed Matter* **2006,** *18* (28), S1151-S1176.
39. Cevc, G.; Schätzlein, A. G.; Richardsen, H.; Vierl, U., Overcoming Semipermeable Barriers, Such as the Skin, with Ultradeformable Mixed Lipid Vesicles, Transfersomes, Liposomes, or Mixed Lipid Micelles. *Langmuir : the ACS journal of surfaces and colloids* **2003,** *19* (26), 10753-10763.
40. Mateos-Maroto, A.; Guerrero-Martínez, A.; Rubio, R. G.; Ortega, F.; Martínez-Pedrero, F., Magnetic Biohybrid Vesicles Transported by an Internal Propulsion Mechanism. *ACS Applied Materials & Interfaces* **2018,** *10* (35), 29367-29377.
41. Chevry, L.; Sampathkumar, N. K.; Cebers, A.; Berret, J. F., Magnetic wire-based sensors for the microrheology of complex fluids. *Physical Review E* **2013,** *88* (6), 062306.
42. Galmarini, M. V.; Baeza, R.; Sanchez, V.; Zamora, M. C.; Chirife, J., Comparison of the viscosity of trehalose and sucrose solutions at various temperatures: Effect of guar gum addition. *LWT - Food Science and Technology* **2011,** *44* (1), 186-190.
43. Frka-Petesic, B.; Erglis, K.; Berret, J. F.; Cebers, A.; Dupuis, V.; Fresnais, J.; Sandre, O.; Perzynski, R., Dynamics of paramagnetic nanostructured rods under rotating field. *Journal of Magnetism and Magnetic Materials* **2011,** *323* (10), 1309-1313.
44. Blake, J. R.; Chwang, A. T., Fundamental singularities of viscous flow. *Journal of Engineering Mathematics* **1974,** *8* (1), 23-29.
45. Lee, S. H.; van Noort, D.; Lee, J. Y.; Zhang, B.-T.; Park, T. H., Effective mixing in a microfluidic chip using magnetic particles. *Lab on a Chip* **2009,** *9* (3), 479-482.
46. Behkam, B.; Sitti, M., Bacterial flagella-based propulsion and on/off motion control of microscale objects. *Appl. Phys. Lett.* **2007,** *90* (2), 023902.
47. Park, S. J.; Park, S.-H.; Cho, S.; Kim, D.-M.; Lee, Y.; Ko, S. Y.; Hong, Y.; Choy, H. E.; Min, J.-J.; Park, J.-O.; Park, S., New paradigm for tumor theranostic methodology using bacteria-based microrobot. *Scientific Reports* **2013,** *3*, 3394.
48. Gosse, C.; Croquette, V., Magnetic Tweezers: Micromanipulation and Force Measurement at the Molecular Level. *Biophysical Journal* **2002,** *82* (6), 3314-3329.
49. Abbott, J. J.; Peyer, K. E.; Lagomarsino, M. C.; Zhang, L.; Dong, L.; Kaliakatsos, I. K.; Nelson, B. J., How Should Microrobots Swim? *The International Journal of Robotics Research* **2009,** *28* (11-12), 1434-1447.

50. Goldman, A. J.; Cox, R. G.; Brenner, H., Slow viscous motion of a sphere parallel to a plane wall—I Motion through a quiescent fluid. *Chemical Engineering Science* **1967,** *22* (4), 637-651.
51. Karatekin, E.; Sandre, O.; Guitouni, H.; Borghi, N.; Puech, P.-H.; Brochard-Wyart, F., Cascades of transient pores in giant vesicles: line tension and transport. *Biophysical journal* **2003,** *84* (3), 1734-1749.
52. Sandre, O.; Moreaux, L.; Brochard-Wyart, F., Dynamics of transient pores in stretched vesicles. *Proc. Natl. Acad. Sci. U. S. A.* **1999,** *96* (19), 10591-10596.
53. Tamba, Y.; Tanaka, T.; Yahagi, T.; Yamashita, Y.; Yamazaki, M., Stability of giant unilamellar vesicles and large unilamellar vesicles of liquid-ordered phase membranes in the presence of Triton X-100. *Biochimica et Biophysica Acta (BBA) - Biomembranes* **2004,** *1667* (1), 1-6.
54. Touitou, E.; Dayan, N.; Bergelson, L.; Godin, B.; Eliaz, M., Ethosomes — novel vesicular carriers for enhanced delivery: characterization and skin penetration properties. *Journal of Controlled Release* **2000,** *65* (3), 403-418.
55. Nishiyama, Y., The mathematics of egg shape. *International Journal of Pure and Applied Mathematics* **2012,** *78* (5), 679-689.
56. Abdallah Daddi-Moussa-Ider, S. G., Benno Liebchen, Christian Hoell, Arnold J. T. M. Mathijssen, Francisca Guzmán-Lastra, Christian Scholz, Andreas M. Menzel, Hartmut Löwen, Membrane penetration and trapping of an active particle. *arXiv:1901.07359* **2019**.
57. Chen, X.-Z.; Hoop, M.; Mushtaq, F.; Siringil, E.; Hu, C.; Nelson, B. J.; Pané, S., Recent developments in magnetically driven micro- and nanorobots. *Applied Materials Today* **2017,** *9*, 37-48.
58. Grosjean, G.; Hubert, M.; Collard, Y.; Pillitteri, S.; Vandewalle, N., Surface swimmers, harnessing the interface to self-propel. *Eur. Phys. J. E* **2018,** *41* (11), 137.
59. Giner-Casares, J. J.; Reguera, J., Directed self-assembly of inorganic nanoparticles at air/liquid interfaces. *Nanoscale* **2016,** *8* (37), 16589-16595.
60. Janssen, X. J. A.; Schellekens, A. J.; van Ommering, K.; van Ijzendoorn, L. J.; Prins, M. W. J., Controlled torque on superparamagnetic beads for functional biosensors. *Biosensors and Bioelectronics* **2009,** *24* (7), 1937-1941.
61. Franke, T.; Schmid, L.; Weitz, D. A.; Wixforth, A., Magneto-mechanical mixing and manipulation of picoliter volumes in vesicles. *Lab on a Chip* **2009,** *9* (19), 2831-2835.
62. Elani, Y.; Law, R. V.; Ces, O., Vesicle-based artificial cells as chemical microreactors with spatially segregated reaction pathways. *Nat. Commun.* **2014,** *5*, 5305.
63. Chiu, D. T.; Wilson, C. F.; Ryttsén, F.; Strömberg, A.; Farre, C.; Karlsson, A.; Nordholm, S.; Gaggar, A.; Modi, B. P.; Moscho, A.; Garza-López, R. A.; Orwar, O.; Zare, R. N., Chemical Transformations in Individual Ultrasmall Biomimetic Containers. *Science* **1999,** *283* (5409), 1892-1895.
64. Walde, P., Enzymatic reactions in liposomes. *Current Opinion in Colloid & Interface Science* **1996,** *1* (5), 638-644.
65. Arcizet, D.; Meier, B.; Sackmann, E.; Rädler, J. O.; Heinrich, D., Temporal Analysis of Active and Passive Transport in Living Cells. *Physical review letters* **2008,** *101* (24), 248103.
66. Kim, K.; Choi, S. Q.; Zasadzinski, J. A.; Squires, T. M., Interfacial microrheology of DPPC monolayers at the air-water interface. *Soft Matter* **2011,** *7* (17), 7782-7789.
67. Angelova, M. I., Liposome Electroformation. In *Giant Vesicles: Perspectives in Supramolecular Chemistry.*, Walde, P. L. L. a. P., Ed. John Wiley and Sons: Chichester, 2000; Vol. 6, pp 27–36.
68. Breton, M.; Amirkavei, M.; Mir, L. M., Optimization of the Electroformation of Giant Unilamellar Vesicles (GUVs) with Unsaturated Phospholipids. *The Journal of Membrane Biology* **2015,** *248* (5), 827-835.
69. Morales-Penningston, N. F.; Wu, J.; Farkas, E. R.; Goh, S. L.; Konyakhina, T. M.; Zheng, J. Y.; Webb, W. W.; Feigenson, G. W., GUV Preparation and Imaging: Minimizing artifacts. *Biochimica et biophysica acta* **2010,** *1798* (7), 1324-1332.

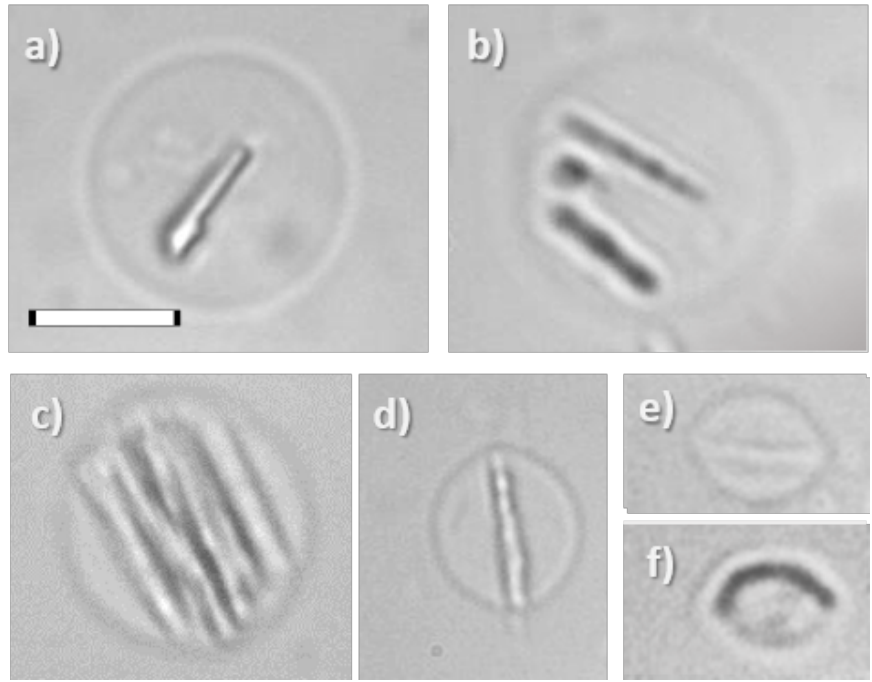

**Figure 1:** Bright-field microscope images showing different unilamellar vesicles encapsulating one or more magnetic nanowires: vesicle of 18 μm in size containing one (a), three (b) or more than five nanowires (c); vesicles of 10 μm and 8 μm, (d) and (e) respectively, slightly deformed by confined wires of 11 and 10 μm; eye-shaped vesicle of 6 μm slightly deformed by a confined and bent nanowire of 9 μm, which manifests as a dark arc in the micrograph (f). Scale bar, 10 μm.

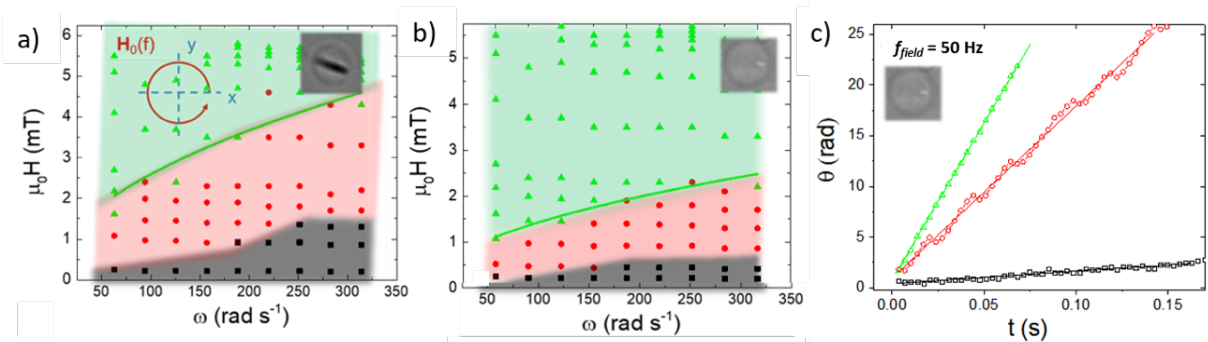

**Figure 2:** State diagrams of two vesicles of 11.2 and 13.1 μm, (a) and (b) respectively, encapsulating nanowires of 11.1 and 3.0 μm in length. By changing the strength and frequency of the applied field the nanowires present synchronous (green triangles) and asynchronous (red circles) rotation. Nanowires do not rotate in the region of low strength (black squares). (c) Time evolution of the angle $\theta$ travelled by the wire confined in the hybrid vesicle described in (b), when the applied field rotates at 314 rad s, and $\mu_0 H = 4$ mT (green triangles), $\mu_0 H = 1.7$ mT (red circumferences) and $\mu_0 H = 0.5$ mT (black squares). Here, the straight lines represent the average angular velocity in each condition.

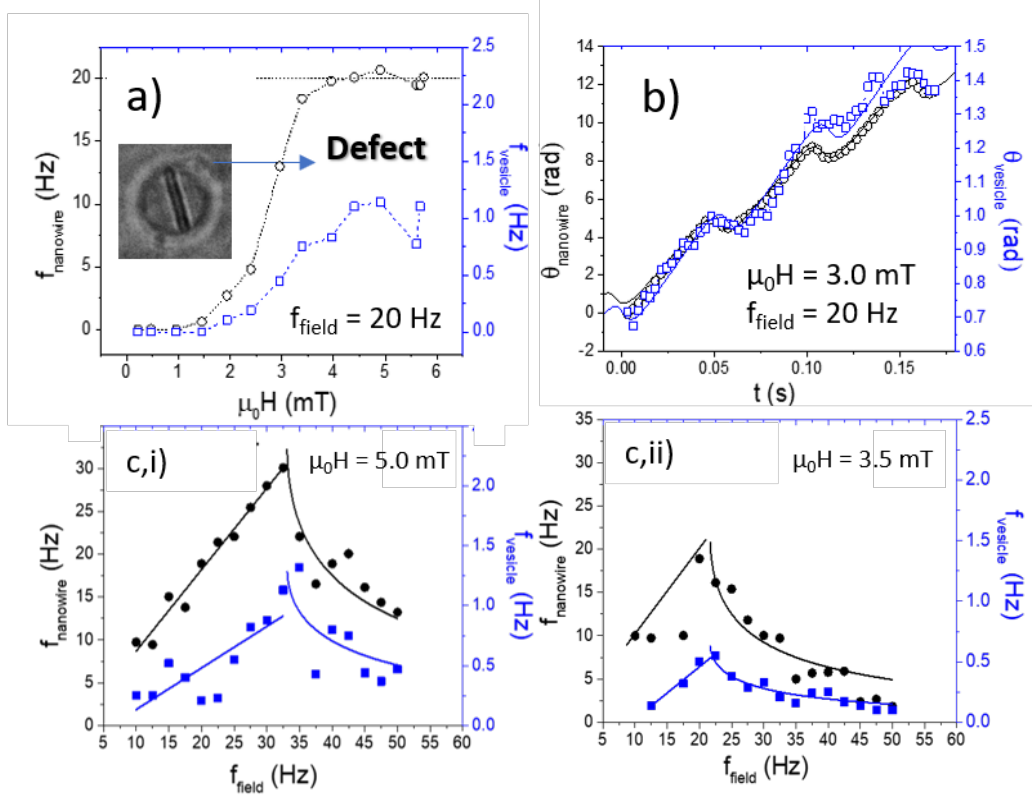

**Figure 3:** (a) Frequency of the rotating nanowire (circles) and membrane (squares) versus the strength of the applied field for a wire 10.5 μm in length encapsulated in a vesicle 11.0 μm in size. The field frequency is $f_{field}$ = 20.0 Hz. (b) Temporal evolution of the angle travelled by the wire (circles) and the vesicle (squares), for $\mu_0 H$ = 3.0 mT and $f_{field}$ = 20.0 Hz. The fitting curves, with a unique fitting parameter $f_c$ = 17.8 Hz, are given by Equations 1 and 2. (c) Frequency of the rotating nanowire (circles) and membrane (squares) versus the frequency of the applied field for a wire 10.5 μm in length encapsulated in a vesicle 11.0 μm in size. The field strength is $\mu_0 H$ = 5.0 mT (i) and $\mu_0 H$ = 3.5 mT (ii). The continues lines represent the trends expected in the synchronous and asynchronous regimes.

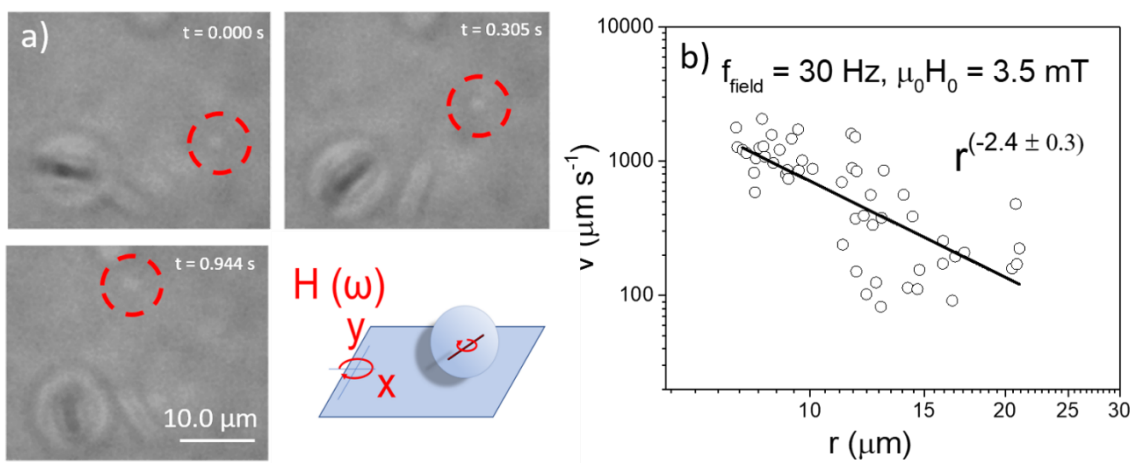

**Figure 4:** (a) The sequence of images shows the trajectory of a small liposome or phospholipid residue advected by the flow generated around the rotating hybrid system. b) The intensity of the flow field $v$ decreases with the distance $r$ to the vesicle center according to $r^{-2.4}$. The nanowire, 10.2 μm in length, is confined in a vesicle 10.5 μm in size. The field, $f_{field}$ = 30.0 Hz and $\mu_0 H_0$ = 3.5 mT, is applied parallel to the substrate.

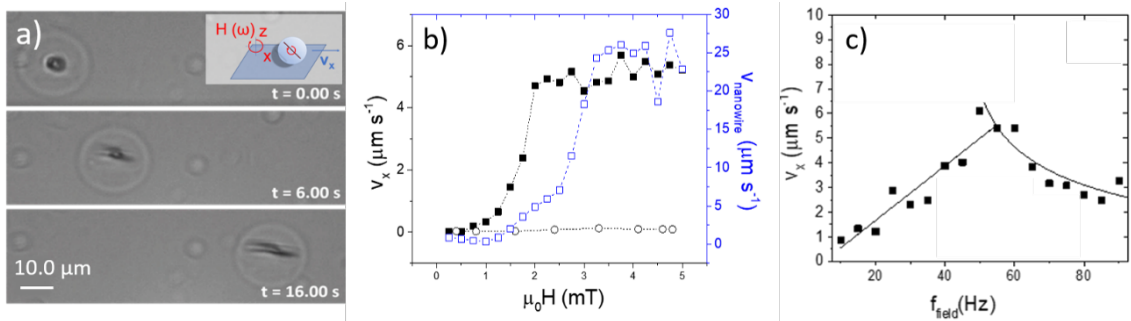

**Figure 5:** (a) The sequence of images shows the translational motion of a hybrid capsule, composed by a rotating nanowire 17.5 μm in length encapsulated in a vesicle 22.8 μm in size, propelling near the glass substrate. The circularly polarized magnetic field, $f_{field}$ = 20.0 Hz and $\mu_0 H$ = 4.0 mT, rotates in the plane perpendicular to the solid surface. (b) Average speed along the $x$ axis of the previous hybrid system (filled squares), $v_x$, of a vesicle 15.9 μm in size loaded with a spherical superparamagnetic particle 2.8 μm in size (empty circles) and a non-encapsulated nanowire 17 μm in length (empty squares), $v_{nanowire}$, versus the rotating field strength. The field frequency is $f_{field}$ = 20.0 Hz. (c) Average speed along the $x$ axis, $v_x$, as a function of the field frequency. Here, the hybrid system is composed by a rotating nanowire 7.2 μm in length encapsulated in a vesicle 11.2 μm in size. The field strength is $\mu_0 H$ = 3.5 mT, and the continues lines fit the trends measured in the synchronous and asynchronous regimes, for $f_c$ = 50 Hz, $D_{vesicle}$ = 11.2 μm, and $\gamma$ = 0.448.

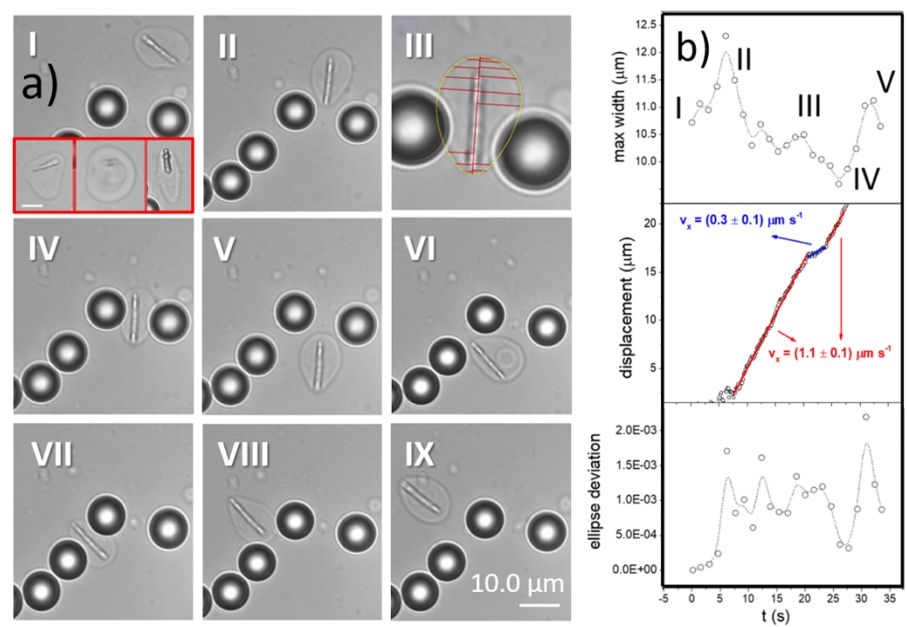

**Figure 6:** (a) The sequence of images shows the steered motion of a deformable hybrid vesicle, dynamically guided through a set of immobile 10-μm polystyrene particles firmly attached to the substrate. The inset in picture I shows different examples of the deformable vesicles observed after the addition of Triton X-100 (0.38 mM) (scale bar: 10 μm). (b) Maximum width, area and ellipse deviation of the vesicle measured when the hybrid vesicle passes between two obstacles (Images I-V, t = 0-35 s).

The realization of microscopic vesicles capable to transport picoliter droplets in a viscous medium has many implications for the delivery of cargos in lab-on-a-chip devices. In this work, vesicles containing magnetic nanowires combine the transport at will of microscopic systems, the protection conferred by the phospholipid membrane and the required flexibility to deform during passage through microscale openings.

**Keyword** magnetic carriers

A. Mateos-Maroto, F. Ortega, R.G. Rubio, J.F. Berret, F. Martínez-Pedrero *
**Title** Giant Vesicles with Encapsulated Magnetic Nanowires as Versatile Carriers, Transported via Rotating and Non-Homogeneous Magnetic Fields

ToC figure

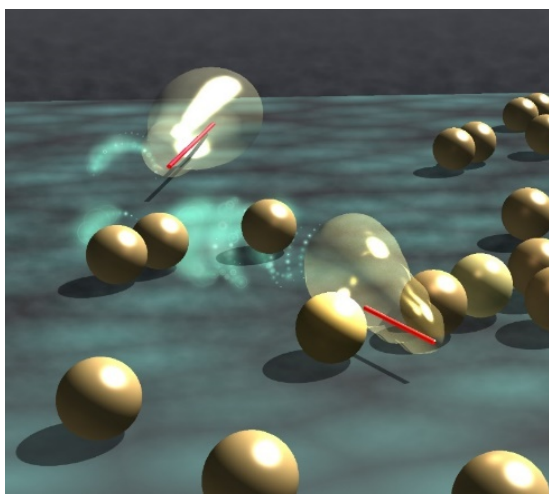